\documentclass[preprint,12pt]{elsarticle}

\usepackage[%
    pdftex,%
    colorlinks=true,%
    hyperindex,%
    plainpages=false,%
    pagebackref=false,%
    bookmarksopen,%
    bookmarksnumbered %
      ]{hyperref}

\usepackage[T1]{fontenc} 
\usepackage[utf8]{inputenc}
\usepackage{amsfonts}
\usepackage{amsthm}
\usepackage[]{algorithm2e}
\usepackage{mathtools}
\usepackage{axodraw4j}
\usepackage{listings}
\usepackage{lipsum}
\usepackage{bold-extra}

\usepackage{amssymb}
\usepackage{tikz}
\usetikzlibrary{arrows,positioning, calc}

\biboptions{sort&compress}

\newcommand*\justify{%
  \fontdimen2\font=0.4em
  \fontdimen3\font=0.2em
  \fontdimen4\font=0.1em
  \fontdimen7\font=0.1em
  \hyphenchar\font=`\-
}

\newcommand*{\defeq}{=} 

\SetAlCapSkip{1em}

\newcommand*{\varset}{\epsilon, \{x_j\}} 
\newcommand*{\invariants}{\{x_j\}} 
\newtheorem{claim}{Claim}
\def\eq#1{Eq.~\eqref{#1}}

\newcounter{bla}

\journal{Computer Physics Communications}

\begin{document}
\begin{frontmatter}



\title{Algorithmic transformation of multi-loop master integrals to a canonical basis with \textit{CANONICA}}

\author{Christoph Meyer}
\cortext[author]{\textit{E-mail address:} christoph.meyer@physik.hu-berlin.de}
\address{Institut f\"ur Physik, Humboldt-Universit\"at zu Berlin, 
    12489 Berlin, Germany\\}

\begin{abstract}
The integration of differential equations of Feynman integrals can be greatly facilitated by using a canonical basis. This paper presents the Mathematica package \textit{CANONICA}, which implements a recently developed algorithm to automatize the transformation to a canonical basis. This represents the first publicly available implementation suitable for differential equations depending on multiple scales. In addition to the presentation of the package, this paper extends the description of some aspects of the algorithm, including a proof of the uniqueness of canonical forms up to constant transformations.
%
%

\end{abstract}


\end{frontmatter}

\begin{picture}(0,0)
\put(300,400){HU-EP-17/10}
\end{picture}

{\bf PROGRAM SUMMARY}\\
\begin{small}
\noindent
{\em Program Title:} \textit{CANONICA}\\
{\em Licensing provisions:} GNU General Public License version 3\\
{\em Programming language:} Wolfram Mathematica, version 10 or higher \\
{\em Nature of problem:} Computation of a rational basis transformation of master integrals leading to a canonical form of the corresponding differential equation.\\
{\em Solution method:} The transformation law is expanded in the dimensional regulator. The resulting differential equations for the expansion coefficients of the transformation are solved with a rational ansatz.\\
%
\end{small}
\section{Introduction}
The calculation of higher order corrections to the cross-sections measured at the LHC is crucial in order to improve the understanding of both the background reactions as well as the signal processes. The current state of the art are NNLO QCD corrections to $2\rightarrow 2$ processes involving a limited number of mass scales. A major challenge in these computations is the evaluation of the occurring Feynman integrals. While the calculation of Feynman integrals can be attempted with numerous approaches, the method of differential equations \cite{Kotikov:1990kg, Remiddi:1997ny, Gehrmann:1999as} has been particularly successful in the recent years \cite{Henn:2013pwa, Henn:2013fah, Henn:2013woa, Henn:2013nsa, Argeri:2014qva, Henn:2014lfa, Caron-Huot:2014lda, Gehrmann:2014bfa, Caola:2014lpa, Li:2014bfa, Hoschele:2014qsa, DiVita:2014pza, vonManteuffel:2014mva, Grozin:2014hna, Bell:2014zya, Huber:2015bva, Gehrmann:2015ora, Gehrmann:2015dua, Bonciani:2015eua, Anzai:2015wma, Grozin:2015kna, Gehrmann:2015bfy, Gituliar:2015iyq, Lee:2016htz, Henn:2016men, Bonciani:2016ypc, Eden:2016dir, Lee:2016lvq, Bonciani:2016qxi, Bonetti:2016brm, Henn:2016kjz, Lee:2016ixa, DiVita:2017xlr, Boels:2017skl, Lee:2017mip}. This success is due to the observation \cite{Henn:2013pwa} that the differential equation can be simplified significantly by turning to a so-called canonical basis of master integrals. The differential equation of a canonical basis of master integrals can easily be integrated in terms of iterated integrals such as multiple polylogarithms \cite{Chen:1977oja,Goncharov:1998kja}. 

It is well known that Feynman integrals exist \cite{Caffo:1998du, Laporta:2004rb, Bloch:2013tra, Adams:2014vja, Bloch:2014qca, Adams:2015gva, Bloch:2016izu, Remiddi:2016gno, Adams:2016xah, Bonciani:2016qxi}, which do not evaluate to this class of functions. These integrals generally satisfy differential equations of higher order. The solutions of the homogeneous part of these equations have been shown to be constructible by evaluating unitarity cuts \cite{Primo:2016ebd, Frellesvig:2017aai, Bosma:2017ens, Primo:2017ipr, Harley:2017qut}\footnote{In fact, unitarity cuts have also been used to derive differential equations of Feynman integrals \cite{Zeng:2017ipr}.}. Some integrals, which exceed the class of multiple polylogarithms, have recently been shown to be iterated integrals of modular forms \cite{Adams:2017ejb}. However, the concept of a canonical basis has not yet been extended to integrals of this kind.

The class of Feynman integrals, which do admit a canonical basis, is still large and contains many integrals of phenomenological interest. It is therefore desirable to automate the calculation of these integrals as much as possible. The systematic application \cite{Laporta:2001dd, Lee:2013mka} of integration by parts relations \cite{Tkachov:1981wb, Chetyrkin:1981qh} to reduce all scalar integrals to a finite number of master integrals has been automated in a variety of publicly available tools \cite{Anastasiou:2004vj, Studerus:2009ye, vonManteuffel:2012np, Lee:2012cn, Smirnov:2013dia, Smirnov:2014hma, Larsen:2015ped, Georgoudis:2016wff, Maierhoefer:2017hyi}. This leaves the process of choosing a canonical basis as the next step to be automated. A number of different methods \cite{Henn:2013pwa, Argeri:2014qva, Caron-Huot:2014lda, Gehrmann:2014bfa, Hoschele:2014qsa, Lee:2014ioa, Henn:2014qga, Eden:2016dir, Meyer:2016slj, Adams:2017tga} have been proposed to construct such a basis or, equivalently, the transformation from a given basis to a canonical basis. Until now, only implementations of the algorithm presented in Ref. \cite{Lee:2014ioa} are publicly available \cite{Gituliar:2016vfa, Prausa:2017ltv, Gituliar:2017vzm}. However, this algorithm is restricted to ordinary differential equations, which are not sufficient to describe the full functional dependence of Feynman integrals depending on multiple dimensionless scales. A wide class of phenomenologically relevant integrals is thus not covered. 

This paper aims to overcome this restriction by introducing an implementation of the algorithm in Ref. \cite{Meyer:2016slj}, which is applicable to multi-scale problems. The accompanying \texttt{\justify Mathematica} package \textit{CANONICA} allows to calculate a rational transformation to a canonical basis for a given differential equation. In addition, the package provides some supplemental functionality for handling differential equations of Feynman integrals. 

The description of the algorithm in Ref. \cite{Meyer:2016slj} is extended in the present paper by a detailed account of the construction of the set of rational functions used for the ansatz. Moreover, the occurrence of non-linear polynomial equations in the parameters of the ansatz is addressed with a procedure to extract all relevant information by solving only linear equations, while maintaining all of the algorithms generality. The latter relies on the uniqueness of canonical forms up to constant transformations. While it is trivial to show that a constant transformation of a given canonical form leads again to a canonical form, it is not obvious that all possible canonical forms can be obtained in this way, which will be proven in this paper.

The paper is organized as follows. In Section \ref{sec:Algorithm} the algorithm in Ref. \cite{Meyer:2016slj} is briefly reviewed and a description of the procedure for the generation of the ansatz is given. Furthermore, the uniqueness of canonical forms up to constant transformations is proven in this section. Building on this result, the treatment of non-linear polynomial equations in the parameters of the ansatz is discussed. Section \ref{sec:package} introduces the \textit{CANONICA} package by outlining the installation and the contents of the package, which is followed by a few examples illustrating the usage of the main features of \textit{CANONICA}. Furthermore, an overview over the hierarchy of the main public functions is given. The conclusions are drawn in Section \ref{sec:Conclusion}. A brief description of all functions and options provided by the package is contained in \ref{App:ListFunctions} and \ref{App:ListOptions}. The global variables and protected symbols of the package are listed in \ref{App:GlobalVars}.

\section{Algorithm}
\label{sec:Algorithm}
This section briefly reviews the algorithm introduced in Ref. \cite{Meyer:2016slj} and presents more details on some aspects of the algorithm. In particular, the procedure used by \textit{CANONICA} to generate an ansatz is described in detail. Furthermore, canonical forms are proven to be unique up to constant transformations. This result allows to attribute the occurrence of non-linear equations in the parameters of the ansatz precisely to this ambiguity. On this basis, a procedure to calculate the transformation by solving only linear equations is outlined.

\subsection{Preliminaries}
\label{subsec:Preliminaries}
Let $\vec{f}(\varset)$ denote the $m$-dimensional vector of master integrals, which depends on the dimensional regulator $\epsilon$ and a set $\{x_j\}$ of dimensionless invariants. By taking the total derivative of the vector of master integrals with respect to the invariants and expressing the result as a linear combination of master integrals, a coupled system of differential equations is obtained:
\begin{equation}
\label{DEQDifferentialForm}
\textup{d}\vec{f}(\varset)=a(\varset)\vec{f}(\varset),
\end{equation}
with
\begin{equation}
a(\varset)=\sum_{i=1}^Ma_i(\varset)\textup{d}x_i.
\end{equation}
Here the $a_i(\varset)$ denote $m\times m$ matrices of rational functions in the invariants and $\epsilon$. Using the linear independence of the master integrals over the field of rational functions in the invariants and taking the exterior derivative of \eq{DEQDifferentialForm} implies the following integrability condition:
\begin{equation}
\label{diffIntegralityCondition}
\textup{d}a-a\wedge a=0,
\end{equation}
which is a valuable consistency check for differential equations in several variables. Transforming the basis of master integrals with an invertible transformation $T$, 
\begin{equation}
\vec{f}=T(\varset)\vec{f}^\prime,
\end{equation}
as suggested in Ref. \cite{Henn:2013pwa}, leads to the following transformation law for $a(\varset)$:
\begin{equation}
\label{DEQTrafo}
a^\prime = T^{-1}aT-T^{-1}\textup{d}T.
\end{equation}
It has been observed  \cite{Henn:2013pwa} that with an appropriate change of the basis of master integrals, the differential equation can often be cast in the following form:
\begin{equation}
\label{EpsForm}
a^\prime(\varset)=\epsilon\textup{d}\tilde{A}(\varset)=\epsilon\sum_{l=1}^N\tilde{A}_l\textup{d}\log(L_l(\invariants)),
\end{equation}
where the $\tilde{A}_l$ denote constant $m\times m$ matrices and the functions $L_l(\invariants)$ are called \textit{letters}. The above form of the differential equation is called \textit{canonical}- or \textit{$\epsilon$-form}. In this form, the integration of the differential equation in terms of iterated integrals is reduced to a merely combinatorial task (c.f. e.g., \cite{Henn:2013woa, Henn:2014lfa, Caron-Huot:2014lda, Gehrmann:2014bfa, Caola:2014lpa}).

\subsection{Review of the algorithm}
\label{subsec:Review}
In this section the algorithm presented in Ref. \cite{Meyer:2016slj} is briefly reviewed. Throughout this section, the existence of a rational transformation of the differential equation into $\epsilon$-form is assumed. The purpose of the algorithm is to compute this transformation for a given differential equation, provided it exists. Any transformation to a canonical basis has to satisfy the following equation:
\begin{equation}
\label{aprimeINeps}
\epsilon\textup{d}\tilde{A}=T^{-1}aT-T^{-1}\textup{d}T,
\end{equation}
for some $\textup{d}\tilde{A}$ of the form in \eq{EpsForm}. The resulting differential form $\textup{d}\tilde{A}$ is unknown and thus has to be determined as well. In Ref. \cite{Meyer:2016slj} it has been proven that the determinant of the transformation is fixed up to a rational function $F(\epsilon)$ by the trace of the differential form $a$
\begin{equation}
\label{trDlogdecomp}
\textup{Tr}[a]=\epsilon X(\{x_j\})+Y(\varset),
\end{equation}
where $X(\{x_j\})$ denotes the sum of dlog-terms with coefficients proportional to $\epsilon$ and $Y(\varset)$ denotes the sum of dlog-terms with constant coefficient. Then the determinant is given by
\begin{equation}
\label{DetIsFixed}
\det(T)=F(\epsilon)\exp\left(\int_\gamma Y(\varset)\right),
\end{equation}
and the trace of the resulting $\epsilon$-form is determined by
\begin{equation}
\label{TraceIsFixed}
\textup{Tr}[\textup{d}\tilde{A}]=X(\invariants).
\end{equation}
For invertible transformations $T$, \eq{aprimeINeps} can equivalently be written as
\begin{equation}
\label{DEQTrafoAlternativ}
\textup{d}T-aT+\epsilon T\textup{d}\tilde{A}=0.
\end{equation}
The basic idea of the algorithm is to expand this equation in $\epsilon$ and solve for the expansion coefficients of the transformation with a rational ansatz. However, the expansion of $T$ may not be finite and therefore an additional step has to be taken, which is reviewed in the following. As $a(\varset)$ is required to be rational in both the invariants and $\epsilon$, a polynomial $h(\varset)$ exists such that $\hat{a}\defeq ah$ has a finite Taylor expansion in $\epsilon$
\begin{equation}
\label{aStartsAtZero}
\hat{a}=\sum_{k=0}^{k_\textup{max}}\epsilon^k\hat{a}^{(k)}.
\end{equation}
In order to fix $h(\varset)$ up to an irrelevant constant factor, $h(\varset)$ is required to satisfy the above condition with the smallest possible number of irreducible factors. The expansion of $h$ is denoted as follows:
\begin{equation}
h(\varset)=\sum_{l=l_\textup{min}}^{l_\textup{max}}\epsilon^l h^{(l)}(\invariants),\quad l_\textup{min}\geq 0.
\end{equation}
Rewriting \eq{DEQTrafoAlternativ} in terms of $\hat{T}=Th$ yields the following equation:
\begin{equation}
\label{DEQfiniteh}
-\hat{T}\textup{d}h+h\textup{d}\hat{T}-\hat{a}\hat{T}+\epsilon h\hat{T}\textup{d}\tilde{A}=0.
\end{equation}
It has been shown in Ref. \cite{Meyer:2016slj} that \eq{DEQfiniteh} will have a solution for $\hat{T}$ with finite expansion, if \eq{DEQTrafoAlternativ} has a rational solution for $T$. This allows to expand $\hat{T}$ in $\epsilon$
\begin{equation}
\label{ThatExpansion}
\hat{T}=\sum_{n=l_\textup{min}}^{n_\textup{max}}\epsilon^n\hat{T}^{(n)},
\end{equation}
and solve \eq{DEQfiniteh} order by order in $\epsilon$ for finitely many coefficients $\hat{T}^{(n)}$. The equations at each order are solved by making an ansatz for $\hat{T}^{(n)}$ in terms of rational functions of the invariants $r_k(\invariants)$
\begin{equation}
\label{ansatzThat}
\hat{T}^{(n)}=\sum_{k=1}^{|\mathcal{R}_T|}\tau_k^{(n)}r_k(\invariants),
\end{equation}
\begin{equation}
\mathcal{R}_T\defeq \left\{r_1(\invariants),\dots,r_{|\mathcal{R}_T|}(\invariants)\right\},
\end{equation}
where the $\tau_k^{(n)}$ denote unknown $m \times m$ matrices independent of the invariants and the regulator, which are to be determined by the algorithm. More details on the choice of the set of rational functions $\mathcal{R}_T$ are given in Section \ref{subsec:AnsatzDB}. For the unknown $\tilde{A}$ an ansatz of the following form can be used:
\begin{equation}
\label{ansatzAtilde}
\tilde{A}=\sum_{l=1}^N\alpha_l \log(L_l(\invariants)),
\end{equation}
where the $\alpha_l$ are considered to be unknown $m\times m$ matrices independent of the invariants and the regulator. The set of polynomials $L_l(\invariants)$ is taken to be the set of irreducible denominator factors of the differential form $a(\varset)$ with trivial dependence on the regulator. In Section \ref{subsec:Ansatzepsform}, this set is shown to contain all letters of the resulting canonical form.

Inserting the ansatz in the expansion of \eq{DEQfiniteh} and requiring the resulting equations to hold for all non-singular values of the invariants implies polynomial equations in the parameters of the ansatz. For more details on the solution of these equations, see Section \ref{subsec:NLequations}.

It is well known that the differential form $a(\varset)$ can be cast in a block-triangular form. This allows to split the computation of the transformation into a recursion over the sectors of the differential equation, which leads to significant performance improvements. With regard to the recursion step, consider a differential equation where all previous sectors have already been transformed into $\epsilon$-form. The first part of the recursion step is to transform the diagonal block of the next sector into $\epsilon$-form with the part of the algorithm described above. After this step, the differential equation is in the following form:
\begin{equation}
\label{aBlockinEpsForm}
a_{I}=\left(\,
\begin{array}{|ccc|c|}
\cline{1-4}
 & & &\\
\quad &   \epsilon\tilde{c} & \quad &\,\, 0 \,\,\\
 & & &\\ \hline
 & b & & \epsilon\tilde{e}\\ \hline
\end{array}
\,\right),
\end{equation}
where $\tilde{c}$ and $\tilde{e}$ are in dlog-form. It has been shown in Ref. \cite{Meyer:2016slj} that the transformation of a differential equation in this form can be split into two parts. First, the off-diagonal block $b$ is transformed into dlog-form with a transformation of the form
\begin{equation}
\label{trafotD}
t_D=\left(\,
\begin{array}{|ccc|c|}
\cline{1-4}
 & & &\\
\quad &   \mathbb{I} & \quad &\,\, 0 \,\,\\
 & & &\\ \hline
 & D & &\mathbb{I}\\ \hline
\end{array}
\,\right),
\end{equation}
which is determined by a differential equation for $D$
\begin{equation}
\label{DDEQ}
\textup{d}D-\epsilon(\tilde{e}D-D\tilde{c})=b-b^\prime.
\end{equation}
Here $b^\prime$ is an unknown quantity, which is required to be in dlog-form. The above equation is solved by first multiplying $D$ by appropriate factors to render its expansion finite. The expansion coefficients are then determined by making an ansatz in terms of a set of rational functions $\mathcal{R}_D$. For more details, see \cite{Meyer:2016slj} and Section \ref{subsec:AnsatzOD}, which describes how the set $\mathcal{R}_D$ is constructed.

The last part of the recursion step is to employ a procedure proposed in Ref. \cite{Lee:2014ioa} to compute a rational transformation in the regulator, which transforms the full differential equation into $\epsilon$-form.

\subsection{Ansatz for diagonal blocks}
\label{subsec:AnsatzDB}
In this section, the choice of the set $\mathcal{R}_T$ of rational functions in the ansatz is discussed. The basic compromise with the ansatz is to choose it large enough to encompass the solution and as small as possible in order to keep the resulting number of equations small and therefore allow the algorithm to perform well. The goal of this section is to present a procedure to generate a finite set of rational functions for a given differential form $a(\varset)$, which can then be used as an ansatz. The first step towards this goal is to determine the set of possible denominator factors of the transformation to a canonical basis. A natural guess is to consider the set of irreducible denominator factors of $\hat{a}$, which is proven in the following to contain all possible factors.

It is useful to first define some notation. Let $f(\invariants)$ be an irreducible polynomial and $S(\varset)$ some matrix-valued rational differential form or function. Then, the notation 
\begin{equation}
S\sim\frac{1}{f^n}
\end{equation}
indicates\footnote{Throughout this paper, the number 0 is understood to be included in the natural numbers.} $n\in\mathbb{N}$ to be the maximal number for which $S$ can be written as
\begin{equation}
S=R\frac{1}{f^n}.
\end{equation}
Here $R$ is required to be nonzero and not to be the product of $f$ and a quantity, which is finite on the set of all zeros of $f$. The set $\mathcal{I}(S)$ of irreducible denominator factors of $S$ is then given by those factors $f$ with $S\sim1/f^k$ and $k\geq 1$.

\begin{claim}
\label{claim:factorsT}
Each irreducible denominator factor $f(\invariants)$ of a rational solution $\hat{T}$ of \eq{DEQfiniteh} is an irreducible denominator factor of $\hat{a}$.
\end{claim}
The assertion in the claim is equivalent to $\mathcal{I}(\hat{T})\subseteq\mathcal{I}(\hat{a})$, which will be proven by showing that
\begin{equation}
\label{Thatdenfactor}
\hat{T}\sim\frac{1}{f^n}, \quad n\geq 1
\end{equation}
implies 
\begin{equation}
\hat{a}\sim\frac{1}{f^k},\quad k\geq 1.
\end{equation}
To this end, \eq{Thatdenfactor} is assumed to hold. It is instructive to rearrange the terms in \eq{DEQfiniteh}
\begin{equation}
\label{rearrangedTL}
-\hat{T}\textup{d}h+h(\textup{d}\hat{T}+\epsilon \hat{T}\textup{d}\tilde{A})=\hat{a}\hat{T}.
\end{equation}
Since $\textup{d}h$ is polynomial, the term $\hat{T}\textup{d}h$ behaves as
\begin{equation}
\label{Tdhorder}
\hat{T}\textup{d}h\sim \frac{1}{f^k},\quad k\leq n.
\end{equation}
Given \eq{Thatdenfactor}, there must be a lowest order $s\in\mathbb{Z}$ in the $\epsilon$-expansion of $\hat{T}$ with
\begin{equation}
\hat{T}^{(s)}\sim\frac{1}{f^n},
\end{equation}
and consequently
\begin{equation}
\label{thatleqn}
\hat{T}^{(s-1)}\sim\frac{1}{f^k},\quad k<n.
\end{equation}
The derivative raises the power of $f$ by one, which implies
\begin{equation}
\textup{d}\hat{T}^{(s)}\sim\frac{1}{f^{n+1}}.
\end{equation}
Note that $\textup{d}\tilde{A}$ is in dlog-form, and therefore 
\begin{equation}
\label{aTildeleq1}
\textup{d}\tilde{A}\sim\frac{1}{f^k},\quad k\leq 1.
\end{equation}
Taking both \eq{thatleqn} and \eq{aTildeleq1} into account, it follows
\begin{equation}
\hat{T}^{(s-1)}\textup{d}\tilde{A}\sim\frac{1}{f^k},\quad k\leq n.
\end{equation}
This implies for the order $s$ of the expansion of $(\textup{d}\hat{T}+\epsilon \hat{T}\textup{d}\tilde{A})$
\begin{equation}
\textup{d}\hat{T}^{(s)}+\hat{T}^{(s-1)}\textup{d}\tilde{A}\sim\frac{1}{f^{n+1}},
\end{equation}
which also holds for the full expression $(\textup{d}\hat{T}+\epsilon \hat{T}\textup{d}\tilde{A})$. Due to the minimality requirement, $h$ does not contain any irreducible factors independent of $\epsilon$ and therefore the multiplication of the term in brackets with $h$ in \eq{rearrangedTL} cannot cancel any power of $f$, because $f$ is independent of $\epsilon$. Since \eq{Tdhorder} shows that the term $\hat{T}\textup{d}h$ is of lower order in $f$ than $h(\textup{d}\hat{T}+\epsilon \hat{T}\textup{d}\tilde{A})$, the whole left-hand side of \eq{rearrangedTL} is of order $1/f^{n+1}$ and consequently the right-hand side as well:
\begin{equation}
\hat{a}\hat{T}\sim\frac{1}{f^{n+1}}.
\end{equation}
Since $\hat{T}$ is only of order $1/f^n$, it can be concluded  
\begin{equation}
\hat{a}\sim\frac{1}{f^k},\quad k\geq 1,
\end{equation}
which proves the claim. Thus, the ansatz can without loss of generality be restricted to the set 
\begin{equation}
\mathcal{Q}=\left\{\frac{x_1^{p_1}\cdots x_M^{p_M}}{f_1^{q_1}\dots f_U^{q_U}}\,\,\,\Bigg|\,\,\, p_1,\dots, p_M, q_1,\dots, q_U \in \mathbb{N}\right\}
\end{equation}
of rational functions with the denominator factors drawn from the set $\mathcal{I}(\hat{a})=\{f_1,\dots,f_U\}$ of irreducible denominator factors of $\hat{a}$.

As was argued in Ref. \cite{Meyer:2016slj}, rational functions may be decomposed in terms of a class of simpler rational functions, called Leinartas functions \cite{Lei78, 2012arXiv1206.4740R}. Let $\mathcal{L}(\mathcal{Q})$ denote a basis of the $K$-span of $\mathcal{Q}$ in terms of Leinartas functions. While $\mathcal{L}(\mathcal{Q})$ is guaranteed to contain the correct ansatz, it is still an infinite set. Therefore, a constructive procedure is needed to generate a finite subset of $\mathcal{L}(\mathcal{Q})$ for a given $a(\varset)$. This procedure should be inexpensive to compute while yielding a correct ansatz for most practical examples. Since the procedure outlined in the following is not proven to generate a correct ansatz, it is important to be able to systematically enlarge the ansatz in a way that is guaranteed to eventually encompass the solution.

The strategy to define a finite subset of $\mathcal{L}(\mathcal{Q})$ is to set restrictions on the powers of the invariants in the numerator as well as on the powers of the denominator factors.

While the powers of those factors occurring in $a(\varset)$ may be suspected to be a good indicator for the powers in the transformation, the following simple example demonstrates this to be false. Consider the differential equation 
\begin{equation}
a(\epsilon,\{x\})=\left(-\frac{\alpha}{x}+\frac{\epsilon}{x}\right)\textup{d}x,\quad \alpha\in\mathbb{Z},
\end{equation}
which contains the factor $x$ with the negative power one. However, for any given integer $\alpha$, the rational transformation to the canonical form reads
\begin{equation}
T(\epsilon,\{x\})=\frac{1}{x^\alpha}.
\end{equation}
Consequently, the transformation can contain any power of the factor $x$, while the power of the same factor in the differential equation remains fixed. A much better predictor is given by the determinant of the transformation, which in the 1-dimensional example above is identical to the transformation itself and therefore always yields the correct power of the factor $x$. For higher dimensional differential equations, the determinant does not fix the transformation but still carries information on the powers of  the irreducible denominator factors of the transformation. Let the determinant of the transformation read
\begin{equation}
\det(T)=F(\varset)\prod_{i=1}^U f_i^{-\lambda_i}, \quad \lambda_i\in\mathbb{Z},\,\, f_i\in\mathcal{I}(\hat{a}),
\end{equation}
where $F(\varset)$ denotes the product of all irreducible factors with a non-trivial dependence on the regulator. Then, for each factor $f_i$ with $\lambda_i>0$, there has to be a component $T_{jl}$ of the transformation satisfying
\begin{equation}
\label{lowerBoundsfi}
T_{jl}\sim \frac{1}{f_i^k}, \quad k\geq \left\lceil{\frac{\lambda_i}{m}}\right\rceil,
\end{equation}
where $m$ denotes the dimension of the differential equation. Thus, the determinant sets lower bounds on the maximal powers of the denominator factors in the transformation, which have to be taken into account in the construction of the ansatz.

In the following, a finite subset of $\mathcal{Q}$ will be constructed, which then leads to a finite subset of $\mathcal{L}(\mathcal{Q})$ by taking a basis of its $K$-span in terms of Leinartas functions. The powers $\lambda_i$ are used to define a set of denominators:
\begin{equation}
\mathcal{D}(\delta_D)=\left\{\frac{1}{f^{p_{i_1}}_{i_1}\cdots f^{p_{i_M}}_{i_M}} \,\,\Bigg|\,\, f_{i_j}\in\mathcal{I}(\hat{a}), 0\leq p_{i}\leq\Theta(\lambda_{i})\lambda_i+\delta_D, i_j \neq i_k\,\, \text{for}\,\, j \neq k\right\},
\end{equation}
which has been restricted to at most $M$ denominator factors with $M$ denoting the number of invariants. Any higher number of polynomials in $M$ invariants is algebraically dependent and therefore reducible in terms of Leinartas functions with $M$ or less denominator polynomials. The parameter $\delta_D\in\mathbb{N}$ has been introduced to define a way to enlarge the set of $\mathcal{D}(\delta_D)$ systematically. The default value is going to be $\delta_D=0$. The lower bounds in \eq{lowerBoundsfi} are satisfied for all allowed values of $\delta_D$. For the numerators, consider the set of all possible monomials up to a fixed bound on their total degree
\begin{equation}
\mathcal{N}(\delta_N)=\left\{x_1^{\nu_1}\cdots x_M^{\nu_M} \,\,\,\Bigg|\,\,\, \nu_1,\dots,\nu_M\in\mathbb{N},\,\,\, \sum_{i=1}^M\nu_i\leq 3+\delta_N\right\},
\end{equation}
where the parameter $\delta_N$ has been introduced to control the highest total degree of the monomials in $\mathcal{N}(\delta_N)$. For the default value $\delta_N=0$ the highest total degree of the numerator monomials is three. This choice is made based on practical examples and is intended to make the default value $\delta_N=0$ work for most cases and at the same time yield a rather small ansatz. Furthermore, it has proven useful to also include the following sets of monomials:
\begin{equation}
\mathcal{N}_\textup{det}=\left\{\text{numerator monomials of}\,\, \det(T) \right\},
\end{equation}
\begin{equation}
\mathcal{N}_a=\left\{ \text{numerator monomials of the}\,\,  \hat{a}^{(k)}(\{x_j\}) \right\},
\end{equation}
in order to capture the correct ansatz in more cases already with the default value $\delta_N=0$. Usually, the inclusion of $\mathcal{N}_\textup{det}$ and $\mathcal{N}_a$ does not significantly enlarge the ansatz, while making the default value work for more examples. Finally, the ansatz $\mathcal{R}_T$ is obtained by computing a basis of Leinartas functions of the $K$-span of the set of rational functions drawing their numerators and denominators from the sets defined above:
\begin{equation}
\mathcal{R}_T(\delta_D,\delta_N)=\mathcal{L}\left(\left\{ \frac{p}{f}  \,\,\,\Bigg|\,\,\, f\in \mathcal{D}(\delta_D),\,\,\, p\in \mathcal{N}(\delta_N)\cup \mathcal{N}_a\cup\mathcal{N}_\textup{det}  \right\}\right).
\end{equation}
The set $\mathcal{R}_T(\delta_D,\delta_N)$ is finite and contains all elements of $\mathcal{L}(\mathcal{Q})$ necessary to represent the elements of $\mathcal{Q}$ with denominators from $\mathcal{D}(\delta_D)$ and numerators from $\mathcal{N}(\delta_N)$. Therefore, by increasing the values of $\delta_D$ and $\delta_N$, the set $\mathcal{R}_T(\delta_D,\delta_N)$ can be systematically extended to the whole set of $\mathcal{L}(\mathcal{Q})$, which contains the correct ansatz. While the correct ansatz is necessarily contained in $\mathcal{L}(\mathcal{Q})$, the choice of the finite subset $\mathcal{R}_T\subset\mathcal{L}(\mathcal{Q})$ presented here is a heuristic procedure. However, the knowledge of upper bounds on $\delta_D$ and $\delta_N$ would be enough to turn the algorithm into a computable criterion for the existence of a rational transformation transforming a given differential equation into canonical form.

\subsection{Ansatz for the resulting canonical form}
\label{subsec:Ansatzepsform}
The ansatz for the resulting canonical form in \eq{ansatzAtilde} requires the knowledge of a set of polynomials in the invariants that encompasses the set of letters of the resulting canonical form. In this section, these letters will be shown to be a subset of the set $\mathcal{I}(a)$ of irreducible denominator factors of the original differential equation with trivial dependence on the regulator. Consider the transformation law \eq{aprimeINeps}
\begin{equation}
\label{trafoLawbracket}
\epsilon\textup{d}\tilde{A}=T^{-1}(aT-\textup{d}T).
\end{equation}
Since the transformation $T$ is rational, the derivative does not alter the set of its denominator factors $\mathcal{I}(\textup{d}T)=\mathcal{I}(T)$. Claim \ref{claim:factorsT} implies $\mathcal{I}(T)\subseteq\mathcal{I}(a)$ and thus
\begin{equation}
\mathcal{I}(aT-\textup{d}T)\subseteq\mathcal{I}(a).
\end{equation}
The denominator factors of $T^{-1}$ can be deduced by writing the inverse as
\begin{equation}
T^{-1}=\det(T)^{-1}\textup{adj}(T).
\end{equation}
The cofactors in the adjugate of $T$ are a sum of products of components of $T$, which implies 
\begin{equation}
\mathcal{I}(\textup{adj}(T))\subseteq\mathcal{I}(T)\subseteq\mathcal{I}(a).
\end{equation}
Due to \eq{DetIsFixed} and \eq{trDlogdecomp}, the determinant of $T$ can be written in the form
\begin{equation}
\det(T)=F(\varset)\prod_{i=1}^U f_i(\invariants)^{-\lambda_i}, \quad \lambda_i\in\mathbb{Z},\,\,\, f_i\in\mathcal{I}(a),
\end{equation}
which leads to 
\begin{equation}
\mathcal{I}(\det(T)^{-1})\subseteq\mathcal{I}(a).
\end{equation}
Thus, the irreducible denominator factors of the right-hand side of \eq{trafoLawbracket} have been shown to be a subset of $\mathcal{I}(a)$, hence the same holds for those of the left-hand side
\begin{equation}
\mathcal{I}(\textup{d}\tilde{A})\subseteq\mathcal{I}(a).
\end{equation}
Since $\mathcal{I}(\textup{d}\tilde{A})$ is equal to the set of letters of $\tilde{A}$, this allows to restrict the set of polynomials in the ansatz in \eq{ansatzAtilde} to the set $\mathcal{I}(a)$.
\subsection{Ansatz for off-diagonal blocks}
\label{subsec:AnsatzOD}
The transformation $t_D$ in \eq{trafotD}, which transforms the off-diagonal blocks into dlog-form, is determined by \eq{DDEQ}. A rational solution of this equation for $D$ is computed by making a rational ansatz. In this section, the set of rational functions $\mathcal{R}_D$ to be used for the ansatz is constructed. First, the set of irreducible denominator factors of $D$ will be shown to be a subset of the irreducible denominator factors of $b$. Here and in the following, these factors are assumed not to depend on the regulator unless stated otherwise. The argument proceeds similarly to the one in the proof of claim \ref{claim:factorsT}. In this case it will even be possible to derive upper bounds on the powers of the irreducible denominator factors of $D$. In a second step, these global upper bounds will be refined to upper bounds for the individual components of $D$, which reduces the number of rational functions in the ansatz considerably.

The set of possible irreducible denominator factors occurring in a rational solution $D$ of 
\begin{equation}
\label{DDEQcopy}
\textup{d}D-\epsilon(\tilde{e}D-D\tilde{c})=b-b^\prime
\end{equation}
can be determined from the denominator factors of $b$. In order to demonstrate this, assume
\begin{equation}
D\sim\frac{1}{f^n},\quad n\geq 1,
\end{equation}
where the same notation as in Section \ref{subsec:AnsatzDB} is used. Then, there exists a lowest order $s\in\mathbb{Z}$ in the expansion of $D$ with
\begin{equation}
D^{(s)}\sim\frac{1}{f^n}
\end{equation}
and therefore 
\begin{equation}
D^{(s-1)}\sim\frac{1}{f^k},\quad k<n.
\end{equation}
The derivative raises the order of $f$ by one 
\begin{equation}
\textup{d}D^{(s)}\sim\frac{1}{f^{n+1}}.
\end{equation}
Since $\tilde{e}$ and $\tilde{c}$ are in dlog-form and thus at most of order $1/f$, it follows
\begin{equation}
\textup{d}D^{(s)}-(\tilde{e}D^{(s-1)}-D^{(s-1)}\tilde{c})\sim\frac{1}{f^{n+1}},
\end{equation}
which in turn implies the left-hand side and therefore also the right-hand side of \eq{DDEQcopy} to be of order $1/f^{n+1}$
\begin{equation}
b-b^\prime\sim\frac{1}{f^{n+1}}.
\end{equation}
As $b^\prime$ is in dlog-form and consequently at most of order $1/f$, it can be concluded 
\begin{equation}
b\sim\frac{1}{f^{n+1}}.
\end{equation}
This result allows to extract upper bounds on the order of the irreducible denominator factors of $D$ from a given $b$. Let $\mathcal{I}(b)=\{f_1,\dots,f_U\}$ denote the set of irreducible denominator factors of $b$ and $\lambda_i$ the order of the denominator factor $f_i$
\begin{equation}
b\sim\frac{1}{f_i^{\lambda_i}}, \quad i=1,\dots,U.
\end{equation}
According to the argument above, the upper bounds $\mu_i$ 
\begin{equation}
D\sim\frac{1}{f^{k_i}_i}, \quad 0\leq k_i \leq \mu_i, \quad i=1,\dots,U,
\end{equation}
are given by
\begin{equation}
\label{globalUpperBounds}
\mu_i=\lambda_i-1,\quad i=1,\dots, U.
\end{equation}
Rather than using these bounds to make an ansatz, it is beneficial to reduce the combinatorics of the ansatz by refining the above bounds. The idea is to infer bounds on the powers of the denominator factors of individual components of the solution $D$ rather than for all components at once. Assume 
\begin{equation}
D_{ij}\sim\frac{1}{f^n}, \quad n\geq 1,
\end{equation}
and let $s\in\mathbb{Z}$ denote the lowest order in the expansion of $D_{ij}$ with
\begin{equation}
D^{(s)}_{ij}\sim1/f^n.
\end{equation}
The derivative raises the power by one
\begin{equation}
\textup{d}D^{(s)}_{ij}\sim\frac{1}{f^{n+1}}.
\end{equation}
Consider a component of the order $s$ in the expansion of \eq{DDEQcopy}
\begin{equation}
\textup{d}D_{ij}^{(s)}-(\tilde{e}D^{(s-1)}-D^{(s-1)}\tilde{c})_{ij}=b^{(s)}_{ij}-b^{\prime(s)}_{ij}.
\end{equation}
Since $b^\prime$ is in dlog-form, this term cannot cancel the order $1/f^{n+1}$ of the derivative term. Therefore at least one of the following cases must be true:
\\\textbf{case 1:}
\begin{equation}
b^{(s)}_{ij}\sim\frac{1}{f^{k}},\quad k\geq n+1,
\end{equation}
\\\textbf{case 2:}
\begin{equation}
(\tilde{e}D^{(s-1)}-D^{(s-1)}\tilde{c})_{ij}\sim\frac{1}{f^{k}},\quad k\geq n+1.
\end{equation}
In case 2, there has to be at least one index $\alpha$ with either
\begin{equation}
\label{eialphaone}
\tilde{e}_{i\alpha}\sim\frac{1}{f^1}\quad \text{and}\quad D_{\alpha j}^{(s-1)}\sim\frac{1}{f^k},\quad k\geq n
\end{equation}
or 
\begin{equation}
\label{eialphazero}
\tilde{e}_{i\alpha}\sim\frac{1}{f^0}\quad \text{and}\quad D_{\alpha j}^{(s-1)}\sim\frac{1}{f^{k}},\quad k\geq n+1,
\end{equation}
or there exists at least one index $\beta$ with either
\begin{equation}
\label{cbetajone}
\tilde{c}_{\beta j}\sim\frac{1}{f^1}\quad \text{and}\quad D_{i\beta}^{(s-1)}\sim\frac{1}{f^k},\quad k\geq n
\end{equation}
or 
\begin{equation}
\label{cbetajzero}
\tilde{c}_{\beta j}\sim\frac{1}{f^0}\quad \text{and}\quad D_{i\beta}^{(s-1)}\sim\frac{1}{f^{k}},\quad k\geq n+1.
\end{equation}
So far, the assumption $D^{(s)}_{ij}\sim 1/f^n$ has been demonstrated to imply either $b^{(s)}_{ij}\sim 1/f^{k}$ for some $k\geq n+1$ (case 1) or that some other component of $D^{(s-1)}$ is of order $1/f^k$ with $k\geq n$ (case 2). In case 2, the whole argument can be applied again to the respective components of $D^{(s-1)}$. This can be repeated until either the lowest order in the expansion is reached and therefore case 2 is not possible anymore or at some point only case 1 is possible due to the structure of $\tilde{e}$ and $\tilde{c}$. Thus, all possible chains of this argument necessarily end with case 1. Since $\tilde{e}$, $\tilde{c}$ and $b$ are known prior to the computation of $D$, the chains can be followed backwards in order to derive upper bounds on the powers of the denominator factors of the components of $D$. The idea is to consider all chains at once and start at the last step by reversing case 1 for all components of $D$. Using the powers of the denominator factors of $b$
\begin{equation}
b_{ij}\sim\frac{1}{f^{\lambda_{k,ij}}_{k}},
\end{equation}
case 1 is reversed for all components by setting the upper bounds $\mu_{k,ij}$ of $D_{ij}$ on $f_k$
\begin{equation}
D_{ij}\sim\frac{1}{f_k^p},\quad 0\leq p\leq \mu_{k,ij},
\end{equation}
to
\begin{equation}
\mu_{k,ij}=\lambda_{k,ij}-1, \quad \forall k,i,j.
\end{equation}
It can then be deduced from $\tilde{e}$ and $\tilde{c}$ for each component which other components could have implied the current bounds via case 2. For instance, if there exists an $\alpha$ with 
\begin{equation}
\tilde{e}_{i\alpha}\sim\frac{1}{f^1},
\end{equation}
and the current bound for the order in $1/f$ of $D_{\alpha j}$ is $n$, case 2 is reversed by setting the bound on the order of $D_{ij}$ to $n$ as well, unless it is already higher. At each step it is checked for all components $D_{ij}$, whether there is an $\tilde{e}_{i\alpha}$ as in \eq{eialphaone} or \eq{eialphazero} or a $\tilde{c}_{\beta j}$ as in \eq{cbetajone} or \eq{cbetajzero}. If this is the case, the bounds are updated accordingly. This is repeated until the bounds do not change anymore, and therefore they incorporate all possible cases. Algorithm \ref{alg:upperBounds} summarizes the procedure.
\begin{algorithm}[h]
\label{alg:upperBounds}
 \KwIn{$\{\lambda_{k,ij}\}$, $\tilde{e}$, $\tilde{c}$}
 \KwOut{Set of upper bounds $\mu_{k,ij}$ with $D_{ij}\sim 1/f^k$ and $0\leq k\leq \mu_{k,ij}$}
 $\mu_{k,ij}=\lambda_{k,ij}-1$.\\
\Repeat{Bounds $\mu$ do not change anymore}{
\ForEach{$k$, $i$, $j$, $\alpha$, $\beta$}{
\lIf{ $\tilde{e}_{i\alpha}\sim\frac{1}{f^0}$}{$\mu_{k,ij}=\textup{max}\left(\mu_{k,ij},\mu_{k,\alpha j}-1\right)$}
\lIf{ $\tilde{e}_{i\alpha}\sim\frac{1}{f^1}$}{$\mu_{k,ij}=\textup{max}\left(\mu_{k,ij},\mu_{k,\alpha j}\right)$}
\lIf{ $\tilde{c}_{\beta j}\sim\frac{1}{f^0}$}{$\mu_{k,ij}=\textup{max}\left(\mu_{k,ij},\mu_{k,i \beta}-1\right)$}
\lIf{ $\tilde{c}_{\beta j}\sim\frac{1}{f^1}$}{$\mu_{k,ij}=\textup{max}\left(\mu_{k,ij},\mu_{k,i \beta}\right)$}
}
}
\Return $\{\mu_{k,ij}\}$
\caption{Determination of upper bounds on the powers of the denominator factors of the components of $D$.}
\end{algorithm}
Since the values of the bounds $\mu_{k,ij}$ can only increase during each iteration in algorithm \ref{alg:upperBounds} and the overall bounds $\mu_k$ given in \eq{globalUpperBounds} are upper bounds on the bounds of the components
\begin{equation}
\mu_{k,ij}\leq \mu_k,\quad \forall k,i,j,
\end{equation}
it is clear that the algorithm terminates after a finite number of steps. 
Using the bounds computed with Algorithm \ref{alg:upperBounds}, the following sets of rational functions can be defined:
\begin{equation}
\mathcal{R}_{ij}(\delta_N)=\left\{ \frac{p}{f_1^{q_1}\dots f_U^{q_U}} \,\,\,\Bigg|\,\,\,p\in\mathcal{N}(\delta_N),\,0\leq q_k\leq\mu_{k,ij}\,\, \forall k\right\},
\end{equation}
with
\begin{equation}
\mathcal{N}(\delta_N)=\left\{x_1^{\nu_1}\cdots x_M^{\nu_M} \,\,\,\Bigg|\,\,\, \nu_1,\dots,\nu_M\in\mathbb{N},\,\,\, \sum_{i=1}^M\nu_i\leq 3+\delta_N\right\}.
\end{equation}
The above argument shows that for high enough $\delta_N$, the component $D_{ij}$ is an element of the $K$-span of $\mathcal{R}_{ij}(\delta_N)$. For the ansatz, a basis of Leinartas functions of the $K$-span of the union of all $\mathcal{R}_{ij}(\delta_N)$ is taken
\begin{equation}
\mathcal{R}_D(\delta_N)=\mathcal{L}\left(\bigcup_{i,j}\mathcal{R}_{ij}(\delta_N)\right).
\end{equation}
It would be more efficient to make a different ansatz for each component of $D$ using $\mathcal{L}(\mathcal{R}_{ij}(\delta_N))$. However, this functionality will only be included in a future version of \textit{CANONICA}.

\subsection{On the uniqueness of canonical bases}
\label{subsec:uniqueness}
The application of a constant invertible transformation $C$ to a differential equation in canonical form obviously preserves the canonical form
\begin{equation}
a^\prime=\epsilon\sum_{l=1}^N\left(C^{-1}\tilde{A}_lC\right)\textup{d}\log(L_l).
\end{equation}
This raises the question whether \emph{all} canonical forms can be obtained in this way. The following claim shows that indeed every canonical form can be obtained by a constant transformation from any other canonical form. In this sense the canonical form of a given differential equation is unique up to constant transformations. 

\begin{claim}
\label{claim:uniqueC}
Let $a(\varset)$ be a differential equation of Feynman integrals and $T_1(\varset)$ and $T_2(\varset)$ be invertible rational transformations, which transform it into the canonical forms $\epsilon\textup{d}\tilde{A}_1(\{x_j\})$ and $\epsilon\textup{d}\tilde{A}_2(\{x_j\})$ respectively. Then there exists a constant invertible transformation $C$ transforming $\epsilon\textup{d}\tilde{A}_1(\{x_j\})$ into $\epsilon\textup{d}\tilde{A}_2(\{x_j\})$.
\end{claim}
Consider the transformation $T=T^{-1}_1T_2$, which transforms $\epsilon\textup{d}\tilde{A}_1$ to $\epsilon\textup{d}\tilde{A}_2$. First, the transformation $T$ has to be shown to be independent of the invariants. The corresponding transformation law reads
\begin{equation}
\label{TrafoLaweps2eps}
\epsilon\textup{d}\tilde{A}_2=T^{-1}\epsilon\textup{d}\tilde{A}_1T-T^{-1}\textup{d}T.
\end{equation}
It is instructive to rewrite this equation: 
\begin{eqnarray}
\textup{d}T&=&\epsilon\left(\textup{d}\tilde{A}_1T-T\textup{d}\tilde{A}_2\right)\\
&=&\epsilon\sum_{l=1}^N\left(\tilde{A}_{1l}T-T\tilde{A}_{2l}\right)\textup{d}\log(L_l).
\end{eqnarray}
The summation over the letters is meant to run over the union of the sets of letters of the two canonical forms, since it is a priori not clear that they both have exactly the same set of letters. The letters are assumed to be irreducible polynomials and the union is meant to remove all scalar multiples of letters as well. Since the transformation law is invariant under the multiplication of $T$ with any rational function $g(\epsilon)$, the $\epsilon$-expansion of $T$ can be assumed to start at the order $\epsilon^0$. Then the first order in the expansion of the above equation reads
\begin{equation}
\textup{d}T^{(0)}=0
\end{equation}
and therefore $T^{(0)}$ has to be constant. At any order $n>0$ the expansion of the above equation is given by
\begin{equation}
\textup{d}T^{(n)}=\sum_{l=1}^N\left(\tilde{A}_{1l}T^{(n-1)}-T^{(n-1)}\tilde{A}_{2l}\right)\textup{d}\log(L_l).
\end{equation}
Assuming $T^{(n-1)}$ to be constant, this equation can easily be integrated
\begin{equation}
T^{(n)}=\sum_{l=1}^N\left(\tilde{A}_{1l}T^{(n-1)}-T^{(n-1)}\tilde{A}_{2l}\right)\log(L_l)+const.
\end{equation}
Since $T_1$ and $T_2$ are assumed to be rational in $\epsilon$ and the invariants, the same holds for $T$ and therefore the coefficients of its $\epsilon$-expansion have to be rational as well. This implies 
\begin{equation}
\tilde{A}_{1l}T^{(n-1)}-T^{(n-1)}\tilde{A}_{2l}=0, \quad \forall\, l
\end{equation}
and consequently $T^{(n)}$ has to be constant. By induction, these arguments imply that all coefficients of the $\epsilon$-expansion of $T$ are constant and therefore $T=T(\epsilon)$. As $T$ is independent of the invariants, the transformation law \eq{TrafoLaweps2eps} has the form
\begin{equation}
\textup{d}\tilde{A}_2=T(\epsilon)^{-1}\textup{d}\tilde{A}_1T(\epsilon).
\end{equation}
It can be concluded that $T(\epsilon)$ transforms $\textup{d}\tilde{A}_1$ to $\textup{d}\tilde{A}_2$ for all non-singular values of $\epsilon$, because the left-hand side does not depend on $\epsilon$. Upon choosing such a value $\epsilon_0$, a constant invertible transformation $C=T(\epsilon_0)$ is obtained, which concludes the proof of the claim. The same argument also holds for the more general case of an algebraic dependence of $T_1$ and $T_2$ on $\epsilon$ and the invariants. Altogether, canonical forms have been shown to be unique modulo $GL(m,K)$ transformations. 

This result explains the origin of the non-linear parameter equations, which are treated in Section \ref{subsec:NLequations}. Moreover, the above result can be utilized for the comparison of two different canonical forms of the same problem, provided they are expressed in the same set of invariants. In this situation, claim \ref{claim:uniqueC} asserts the existence of a constant transformation relating the two canonical forms. This can be tested by checking whether the following system of linear equations
\begin{equation}
C\tilde{A}_{2l}=\tilde{A}_{1l}C,\quad l=1,\dots,N,
\end{equation}
has a non-singular solution for the components of $C$.

The uniqueness of the canonical form also manifests itself in \eq{DDEQ}, which governs the transformation of the off-diagonal blocks into dlog-form. In the following, the rational solution of this equation is proven to be unique up to the addition of terms depending solely on the regulator. In practice, this result allows to exclude terms with trivial dependence on the invariants from the ansatz without losing generality.

For a given $b$, let $D$ and $b^\prime$ satisfy \eq{DDEQ}
\begin{equation}
\label{DDEQcopy2}
\textup{d}D-\epsilon(\tilde{e}D-D\tilde{c})=b-b^\prime,
\end{equation}
with $b^\prime$ understood to be in dlog-form. Adding a term $C(\epsilon)$ to the solution $\tilde{D}=D+C(\epsilon)$ solves the same equation
\begin{equation}
\textup{d}\tilde{D}-\epsilon(\tilde{e}\tilde{D}-\tilde{D}\tilde{c})=b-\tilde{b}^\prime,
\end{equation}
with
\begin{equation}
\tilde{b}^\prime=b^\prime+\epsilon(\tilde{e}C(\epsilon)-C(\epsilon)\tilde{c}),
\end{equation}
which is also in dlog-form, since $\tilde{e}$ and $\tilde{c}$ are in dlog-form. This argument establishes the freedom to add terms independent of the invariants to a solution of \eq{DDEQ}. The following argument proves this to be the only possible relation between two solutions of \eq{DDEQcopy2}. Let $D_1$ and $D_2$ satisfy \eq{DDEQcopy2} for a given $b$
\begin{eqnarray}
\textup{d}D_1-\epsilon(\tilde{e}D_1-D_1\tilde{c})=b-b_1^\prime,\\
\textup{d}D_2-\epsilon(\tilde{e}D_2-D_2\tilde{c})=b-b_2^\prime.
\end{eqnarray}
Then the difference $\bar{D}=D_1-D_2$ satisfies
\begin{equation}
\textup{d}\bar{D}-\epsilon(\tilde{e}\bar{D}-\bar{D}\tilde{c})=b_2^\prime-b_1^\prime.
\end{equation}
Let $\hat{\bar{D}}=\bar{D}\epsilon^\tau$ be defined such that the expansion of $\hat{\bar{D}}$ starts at the constant order. The equation for $\hat{\bar{D}}$ reads
\begin{equation}
\label{DhatbarDEQ}
\textup{d}\hat{\bar{D}}-\epsilon(\tilde{e}\hat{\bar{D}}-\hat{\bar{D}}\tilde{c})=B,
\end{equation}
with $B=\epsilon^\tau(b_2^\prime-b_1^\prime)$, which is in dlog-form. The first order in the expansion of \eq{DhatbarDEQ} reads
\begin{equation}
\textup{d}\hat{\bar{D}}^{(0)}=\sum_{l=1}^NB_l^{(0)}\textup{d}\log(L_l),
\end{equation}
which integrates to
\begin{equation}	
\hat{\bar{D}}^{(0)}=\sum_{l=1}^NB_l^{(0)}\log(L_l)+const.
\end{equation}
As $D_1$ and $D_2$ are assumed to be rational, $\hat{\bar{D}}$ has to be rational as well and therefore
\begin{equation}
B_l^{(0)}=0,\quad l=1,\dots,N,
\end{equation}
which implies that $\hat{\bar{D}}^{(0)}$ is constant. Consider the expansion of \eq{DhatbarDEQ} at some order $n>0$
\begin{equation}
\textup{d}\hat{\bar{D}}^{(n)}=(\tilde{e}\hat{\bar{D}}^{(n-1)}-\hat{\bar{D}}^{(n-1)}\tilde{c})+\sum_{l=1}^NB_l^{(n)}\textup{d}\log(L_l).
\end{equation}
The right-hand side is in dlog-form for constant $\hat{\bar{D}}^{(n-1)}$ and therefore $\hat{\bar{D}}^{(n)}$ can only be rational if it is constant as well. This proves by induction that $\hat{\bar{D}}$ is independent of the invariants. Consequently, the difference of the solutions $\bar{D}=\hat{\bar{D}}\epsilon^{-\tau}$ has to be independent of the invariants as well. Altogether, the argument establishes the uniqueness of a rational solution for $D$ of \eq{DDEQ} up the addition of terms that are independent of the invariants. This fact can be used in practice to fix this freedom without losing generality.

\subsection{Treatment of non-linear parameter equations}
\label{subsec:NLequations}
In the course of applying the algorithm, the ansatz in \eq{ansatzThat} and \eq{ansatzAtilde} is inserted in the expansion of \eq{DEQfiniteh}. By requiring the resulting equations to hold for all allowed values of the invariants, a system of equations in the unknown parameters is obtained at each order of the expansion. Due to the term $\epsilon T\textup{d}\tilde{A}$ in \eq{DEQTrafoAlternativ}, these equations can be non-linear. Instead of directly solving these non-linear equations, it will be shown in the following how they can be reduced to linear equations by imposing appropriate constraints.

In Section \ref{subsec:uniqueness}, it has been shown that the resulting canonical form is uniquely fixed up to an invertible constant transformation. Exactly this ambiguity leads to the non-linear equations, because if $\textup{d}\tilde{A}$ was fixed, the term $\epsilon T\textup{d}\tilde{A}$ would not generate non-linear equations. Therefore, the non-linear equations can be turned into linear equations by fixing the degrees of freedom in the ansatz corresponding to a subsequent invertible constant transformation. In order to fix these degrees of freedom directly, they would have to be disentangled from those which are determined by the equations in the parameters. Since this would require a parameterization of the solution set of the non-linear equations, which is essentially equivalent to solving them, a more indirect approach of fixing the freedom is taken in the following.

To this end, suppose for the moment that the parameters of the ansatz can be separated in those which are fixed by the parameter  equations $\{\tau\}$ and those which correspond to the remaining freedom $\{\tau^\prime\}$. Let $T(\epsilon, \{x_j\}, \{\tau\},\{\tau^\prime\})$ be a solution of \eq{DEQTrafoAlternativ}, provided the parameters $\{\tau\}$ solve the parameter equations. According to the proof of claim \ref{claim:uniqueC}, this transformation can be thought of as the product of some fixed transformation $T_1(\varset,\{\tau\})$, which transforms the original differential equation to some canonical form $\epsilon\textup{d}\tilde{A}_1$, and a transformation $C(\epsilon,\{\tau^\prime\})$ parameterizing the transformation of $\textup{d}\tilde{A}_1$ to any other possible canonical form $\epsilon\textup{d}\tilde{A}_2(\tau^\prime)$
\begin{equation}
\label{freeParamFactorize}
T(\epsilon, \{x_j\}, \{\tau\},\{\tau^\prime\})=T_1(\varset,\{\tau\})C(\epsilon,\{\tau^\prime\}),
\end{equation}
\begin{equation}
\label{remainingFreedomTrafo}
\textup{d}\tilde{A}_2(\{\tau^\prime\})=C(\epsilon,\{\tau^\prime\})^{-1}\textup{d}\tilde{A}_1C(\epsilon,\{\tau^\prime\}).
\end{equation}
It should be noted that $\tilde{A}_2$ does in general only depend on a subset of the parameters $\{\tau^\prime\}$, because some parameters can correspond to a non-trivial $\epsilon$-dependence of $C(\epsilon,\{\tau^\prime\})$. As mentioned above, the goal is to fix the resulting differential equation $\textup{d}\tilde{A}_2$ by fixing the corresponding parameters of $\{\tau^\prime\}$. 
This can be achieved by demanding $C(\epsilon,\{\tau^\prime\})$ to equal some fixed constant invertible transformation at some non-singular value $\epsilon=\epsilon_0$. Since the left-hand side of \eq{remainingFreedomTrafo} does not depend on $\epsilon$, this completely fixes $\textup{d}\tilde{A}_2$ irrespective of the particular value of $\epsilon_0$. However, fixing $C(\epsilon,\{\tau^\prime\})$ directly would require the computation of the factorization in \eq{freeParamFactorize}, which is only possible if the separation of the parameters into the sets $\{\tau\}$ and $\{\tau^\prime\}$ is known. Instead, $C(\epsilon,\{\tau^\prime\})$ can be fixed indirectly by demanding
\begin{equation}
\label{addconstraints}
T(\epsilon_0,\{x_{0j}\},\{\tau\},\{\tau^\prime\})=\mathbb{I}
\end{equation}
to hold at some non-singular point $\{x\}=\{x_{j0}\}$, $\epsilon=\epsilon_0$. This is equivalent to fixing $C(\epsilon,\{\tau^\prime\})$ as follows:
\begin{equation}
C(\epsilon_0,\{\tau^\prime\})=T_1(\epsilon_0,\{x_{0j}\},\{\tau\})^{-1}.
\end{equation}
The constraints given by \eq{addconstraints} can be imposed without being able to separate the parameters into $\{\tau\}$ and $\{\tau^\prime\}$. Moreover, these constraints are linear in both the $\{\tau\}$ and the $\{\tau^\prime\}$, since the ansatz in \eq{ansatzThat} is linear in all parameters. Therefore, the additional constraints in \eq{addconstraints} can be used to completely fix the resulting canonical form, which turns the non-linear parameter equations into linear equations.

Recall that the parameter equations are generated order by order in the expansion of \eq{DEQfiniteh} and at each order it is tested whether the series terminates at the current order. The constraints in \eq{addconstraints} can only be imposed if the full $T(\epsilon, \{x_j\}, \{\tau\},\{\tau^\prime\})$ is known. Thus, the computation must have reached the order at which the series terminates. However, non-linear equations can already occur at lower orders in the expansion, i.e. before \eq{addconstraints} can be imposed to turn them into linear equations. The strategy described in the following overcomes this point by essentially just solving the linear equations at each order and keeping the non-linear equations until the constraints can be imposed.

At each order in the expansion of the transformation, the linear equations are solved first and then their solution is inserted into the non-linear ones, which possibly turns some of them into linear equations. These newly generated linear equations can again be solved. This procedure is iterated until no further linear equations are generated. The remaining non-linear equations are kept unsolved. It is then tested whether the series terminates at the current order by generating the additional parameter equations corresponding to this assumption (cf. Ref. \cite{Meyer:2016slj}). The linear equations of these additional equations are then iteratively solved as described above, while the previously obtained still unsolved non-linear equations are taken into account as well. If it turns out during this iteration that the system has no solution, the algorithm proceeds with the next order in the expansion of the transformation. If this is not the case and some non-linear equations remain at the end of the iteration, \eq{addconstraints} is imposed. If the series does terminate at the current order, the additional constraints will turn the remaining non-linear equations into linear ones, which then determine the transformation. If either non-linear equations remain or the linear ones have no solution, it can be concluded that the series does not terminate at the current order and the algorithm proceeds with the next order in the expansion.

Altogether, this procedure allows to compute a transformation to a canonical form by only solving linear equations at each order without sacrificing the generality of the algorithm.

\section{The \textit{CANONICA} package}
\label{sec:package}
This section introduces the \textit{CANONICA} package, which implements the algorithm proposed in Ref. \cite{Meyer:2016slj}. After describing the installation and the contents of the package, the main functionality is illustrated with short usage examples. A detailed description of all functions and options of \textit{CANONICA} can be found in the interactive manual \texttt{\justify ./manual.nb}.

\subsection{Installation}
\label{subsec:Installation}
\textit{CANONICA} is a \texttt{\justify Mathematica} package and requires an installation of version 10 or higher of \texttt{\justify Mathematica}. The \textit{CANONICA} repository can be copied to the local directory with
\begin{verse}
\begin{verbatim}
git clone https://github.com/christophmeyer/CANONICA.git
\end{verbatim}
\end{verse}
Alternatively, an archive file can be downloaded at
\begin{verse}
\begin{verbatim}
https://github.com/christophmeyer/CANONICA/archive/v1.0.tar.gz
\end{verbatim}
\end{verse}
which may be extracted with
\begin{verse}
\begin{verbatim}
tar -xvzf CANONICA-1.0.tar.gz
\end{verbatim}
\end{verse}
There is no further installation necessary, in particular, there are no dependencies other than \texttt{\justify Mathematica}. In a \texttt{\justify Mathematica} session, the package can be loaded by 
\begin{verse}
\begin{verbatim}
Get["CANONICA.m"];
\end{verbatim}
\end{verse}
provided the file \texttt{\justify CANONICA.m} is placed either in the current working directory or in one of the search paths. If this is not the case, \texttt{\justify Get} either has to be called with the full path of the file \texttt{\justify CANONICA.m}, or its location has to be added to the list of Mathematica's search paths, which is stored in the global variable \texttt{\justify \$Path}, by e.g. 
\begin{verse}
\begin{verbatim}
AppendTo[$Path,"/path/to/CANONICA/src/"]
\end{verbatim}
\end{verse}
Changes to \texttt{\justify \$Path} can be made permanent by adding them to the initialization file \texttt{\justify init.m}.
\subsection{Files of the package}
\label{subsec:Files}
The root directory of the \textit{CANONICA} package contains the following files and directories.
\begin{description}
\item[\texttt{./src/CANONICA.m}]~\\Contains all of the source code of the program, in particular, all function definitions as well as short usage messages for the public functions and options.
\item[\texttt{./manual.nb}]~\\An interactive manual in the \texttt{\justify Mathematica} notebook format explaining the usage of all functions and options with short examples.
\item[\texttt{./examples}]~\\Several examples are provided in this directory. The directory of each example contains a .m file with the corresponding differential equation and a .nb notebook file illustrating the application of \textit{CANONICA} to this example. The calculation of the full transformation can also be run in terminal mode with the script \texttt{\justify RunExample.m}. The script is started by calling
\begin{verse}
\begin{verbatim}
math -run "<<RunExample.m"
\end{verbatim}
\end{verse}
or
\begin{verse}
\begin{verbatim}
math -script RunExample.m
\end{verbatim}
\end{verse}
Some basic information about the examples, such as the master integrals and the definition of the kinematic invariants, is provided in the \texttt{\justify ./examples/examples.pdf} file.
\item[\texttt{./LICENSE}]~\\A copy of the third version of the GNU General Public License.
\item[\texttt{./README}]~\\A README file providing basic information on the package.
\end{description}

\subsection{Usage examples}
\label{subsec:UsageEx}
In this section, the main features and their usage are illustrated with short examples. A similar, but more extensive account of this functionality can be found in the manual notebook file, which comes with the package.

The most common input required by \textit{CANONICA} is a differential equation of the form in \eq{DEQDifferentialForm}, which is determined by the differential form $a(\varset)$. Consider the following example:
\begin{align}
a(\epsilon,\{x,y\})&=\left(\,
\begin{array}{cc}
-\frac{2+\epsilon}{x} & 0 \\
0 & -\frac{1+\epsilon}{x}  \\
\end{array}
\,\right)\textup{d}x \nonumber \\
&+\left(\,
\begin{array}{cc}
0 & 0 \\
\frac{(-1+\epsilon)x}{(-1+y)y} & \frac{1-\epsilon(1+y)}{(-1+y)y}  \\
\end{array}
\,\right)\textup{d}y,
\end{align}
depending on the invariants $x$ and $y$. The differential form is represented in \textit{CANONICA} as a list of the matrix-valued coefficients of the differentials of the invariants. The dimensional regulator $\epsilon$ has to be denoted by the protected symbol \texttt{\justify eps}. For the above example, the input reads
\begin{verse}
\begin{verbatim}
a = {
  {{-(2+eps)/x, 0}, {0, -(1+eps)/x}}
  ,
   {{0, 0}, {((-1+eps)x)/((-1+y)y), (1-eps(1+y))/((-1+y)y)}}
 };
\end{verbatim}
\end{verse}
The order of the coefficient matrices has to be specified by a list of the corresponding invariants.
\begin{verse}
\begin{verbatim}
invariants = {x, y};
\end{verbatim}
\end{verse}
The algorithm to compute a transformation of a differential equation to canonical form as outlined in Section \ref{subsec:Review} and in Ref. \cite{Meyer:2016slj} is implemented in the function \texttt{\justify TransformDiagonalBlock}. As the name suggests, this function is intended to be used for calculating the transformation of diagonal blocks of differential equations. For the example above, the function is called as follows:
\begin{verse}
\begin{verbatim}
res=TransformDiagonalBlock[a, invariants]
\end{verbatim}
\end{verse}
which returns the output
\begin{verbatim}
      {
       {{(1-2eps)/x^2, (1-2eps)/x^2}, {(1-eps)/x, (1-eps)/(xy)}}
       ,
       {
        {{-(eps/x), 0}, {0, -(eps/x)}}
        ,
        {{-eps/(-1+y),-eps/(y-y^2)}, {eps/(-1+y),eps/(y-y^2)}}
       }
      }
\end{verbatim}
The output of \texttt{\justify TransformDiagonalBlock} is a list with two entries. The first contains the transformation and the second contains the resulting differential equation in $\epsilon$-form. The resulting differential equation is of course redundant, since it can be computed by applying the transformation to the original differential equation. However, the resulting differential equation is generated anyway in the course of the computation of the transformation and applying the transformation can be a costly operation in itself for larger differential equations. 

The application of a transformation to a differential equation, according to the transformation law \eq{DEQTrafo}, is implemented in the function \texttt{\justify TransformDE}, which for some transformation 
\begin{verse}
\begin{verbatim}
trafo = res[[1]]
\end{verbatim}
\end{verse}
is called as
\begin{verse}
\begin{verbatim}
TransformDE[a, invariants, trafo]
\end{verbatim}
\end{verse}
and returns the resulting differential equation. In order to apply \eq{DEQTrafo}, the inverse of the transformation needs to be computed, which can consume significant computation time for larger matrices, when done with the build-in \texttt{\justify Mathematica} command. However, the transformations usually exhibit a block-triangular structure, which is exploited by \texttt{\justify TransformDE} leading to a considerably better performance.

The function \texttt{\justify TransformDiagonalBlock} is in principle applicable to differential equations of any size. However, the performance can be improved significantly by splitting the computation according to the block-triangular structure of the differential equation and performing the computation in a recursion over the sectors of the differential equation. The main function in \textit{CANONICA} for this purpose is \texttt{\justify RecursivelyTransformSectors}. In addition to the two arguments related to the differential equation itself, this function expects an argument that defines the boundaries of the diagonal blocks. The differential equation in the example above actually splits into two blocks of dimension one, and in this case the boundaries read
\begin{verse}
\begin{verbatim}
boundaries = {{1, 1}, {2, 2}};
\end{verbatim}
\end{verse}
Each entry of the boundaries list corresponds to one diagonal block, which is specified by the position of its lowest and highest integral. Instead of using \texttt{\justify TransformDiagonalBlock} to transform \texttt{\justify a} into $\epsilon$-form all at once, the following command  
\begin{verse}
\begin{verbatim}
RecursivelyTransformSectors[a, invariants, boundaries, {1, 2}]
\end{verbatim}
\end{verse}
computes the transformation in a recursion over the sectors, as described in Section \ref{subsec:Review}. The last argument determines the sectors at which the computation starts and ends. The output is of the same format as described above for \texttt{\justify TransformDiagonalBlock}. If some lower sectors have already been transformed into $\epsilon$-form and the computation should therefore not start at the first sector, the differential equation of the lower-sectors in $\epsilon$-form and the transformation leading to it need to be provided as two additional arguments.

\begin{figure}[h!]
\begin{tikzpicture}[thick,scale=0.9, every node/.style={scale=0.9}]
\clip(-4,-2) rectangle (11,8);

\def\frow{0}
\def\srow{7}
\def\mrow{\frow*0.5+\srow*0.5}

\def\vadd{-4}
\def\vvadd{-4}
\def\vvvadd{3}

\small
\tikzset{
bignodeone/.style={rectangle,rounded corners, draw=black, top color=white, inner    sep=1em,minimum width=5.80cm, minimum height=1.2cm, text centered},
regnodeone/.style={rectangle,draw=black, top color=white, inner sep=1pt,minimum width=5.8cm, minimum height=1cm, text centered},
bignode/.style={rectangle,rounded corners, draw=black, top color=white, inner    sep=1em,minimum width=5.80cm, minimum height=3cm, text centered},
regnode/.style={rectangle,draw=black, top color=white, inner sep=1pt,minimum width=5.8cm, minimum height=1cm, text centered},
nregnode/.style={rectangle,draw=white, top color=white, inner sep=1pt,minimum width=1.2cm, minimum height=0.4cm,text width=5cm}
}
\node[bignodeone] (axiif) at (\mrow,3.5+\vvvadd) {};
\node[regnode] (RecursivelyTransformSectors) at (\mrow, 4.5+\vvvadd) {\texttt{RecursivelyTransformSectors}};
\node[nregnode] (pTransformNextSector) at (\mrow,3.6+\vvvadd) {\texttt{TransformNextSector}}; 

\node[bignode] (axiif) at (\mrow,3.5) {};
\node[regnode] (sTransformNextSector) at (\mrow, 4.5) {\texttt{TransformNextSector}};
\node[nregnode] (pTransformNextDiagonalBlock) at (\mrow,3.6) {\texttt{TransformNextDiagonalBlock}}; 
\node[nregnode] (pTransformOffDiagonalBlock) at (\mrow,3.0) {\texttt{TransformOffDiagonalBlock}};
\node[nregnode] (pTransformDlogToEpsForm) at (\mrow,2.4) {\texttt{TransformDlogToEpsForm}};

\node[bignode] (axiif) at (\frow,3.5+\vadd) {};
\node[regnode] (sTransformNextDiagonalBlock) at (\frow, 4.5+\vadd) {\texttt{TransformNextDiagonalBlock}};
\node[nregnode] (pCalculateNexta) at (\frow,3.6+\vadd) {\texttt{CalculateNexta}}; 
\node[nregnode] (pTransformDiagonalBlock) at (\frow,3.0+\vadd) {\texttt{TransformDiagonalBlock}};

\node[bignodeone] (axiif) at (\srow,3.5+\vvadd) {};
\node[regnode] (sTransformOffDiagonalBlock) at (\srow, 4.5+\vvadd) {\texttt{TransformOffDiagonalBlock}};
\node[nregnode] (pCalculateNextSubsectorD) at (\srow,3.6+\vvadd) {\texttt{CalculateNextSubsectorD}}; 

\draw[draw=black,line width=0.5mm,solid, -triangle 90] (\mrow-3,3.6+\vvvadd) .. controls (\mrow-4.5,3.6+\vvvadd-0.3) and (\mrow-4.5,3.6+\vvvadd-1.7) .. (\mrow-3, 4.5);

\draw[draw=black,line width=0.5mm,solid, -triangle 90] (\mrow-3,3.6) .. controls (\mrow-4.5,3.6) and (\frow-6,3.6+\vvvadd-3.6) .. (\frow-3, 4.5+\vadd);

\draw[draw=black,line width=0.5mm,solid, -triangle 90] (\mrow+3,3.0) .. controls (\srow+4.5,3.0) and (\srow+4.5,\vvvadd-1) .. (\srow+3, 4.5+\vadd);

\end{tikzpicture} 
\medskip
\caption{Hierarchy of the main functions in \textit{CANONICA}. Each block lists the public functions called by the function in the blocks title.} 
\label{fig:hierarchy}
\end{figure}
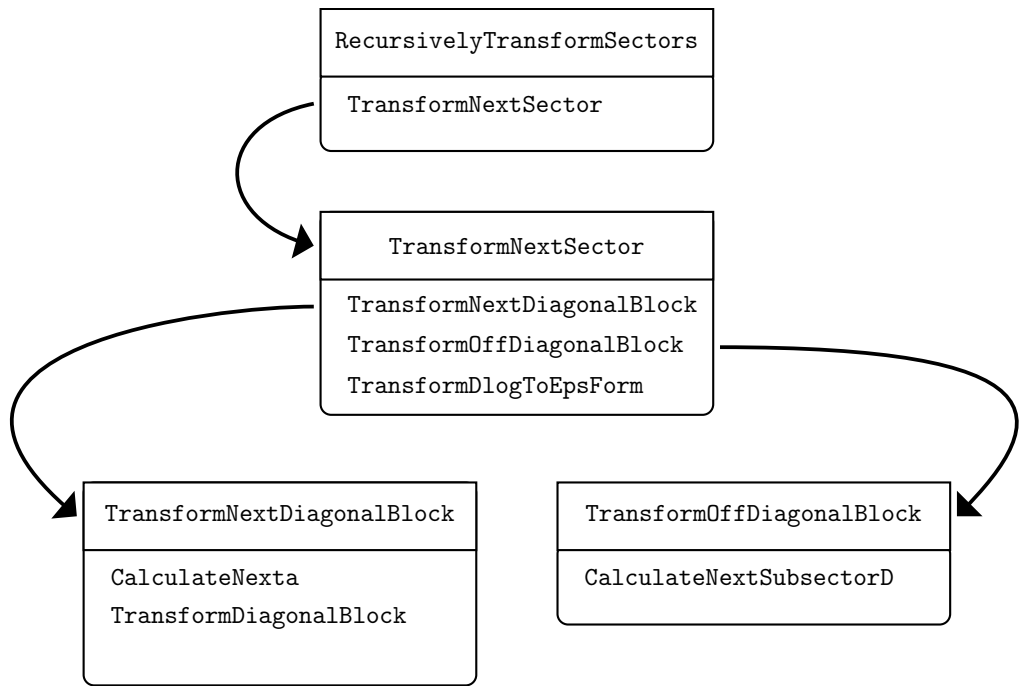

\textit{CANONICA} also has functionality to extract the boundaries of the diagonal blocks from the differential equation. The function \texttt{\justify SectorBoundariesFromDE} extracts the most fine grained boundaries compatible with the differential equation. For instance, in the example above
\begin{verse}
\begin{verbatim}
SectorBoundariesFromDE[a]
\end{verbatim}
\end{verse}
returns 
\begin{verse}
\begin{verbatim}
{{1, 1}, {2, 2}}
\end{verbatim}
\end{verse}
The boundaries obtained in this way may be too fine for the algorithm to find the solution, since the solution space could be constrained too much by splitting the transformation into smaller blocks. It is safer to choose the boundaries according to the sector-ids of the integrals, which in general yields coarser grained boundaries. For a given list of integrals specified by their propagator powers
\begin{verse}
\begin{verbatim}
masterIntegrals={Int["T1", {0, 1, 0, 1, 0, 1, 0, 0, 0}],
                 Int["T1", {0, 1, 0, 1, 1, 1, 0, 0, 0}], 
                 Int["T1", {0, 0, 1, 1, 1, 1, 0, 0, 0}],
                 Int["T1", {0, 1, 1, 1, 1, 1, 0, 0, 0}], 
                 Int["T1", {1, 1, 0, 0, 0, 0, 1, 0, 0}], 
                 Int["T1", {1, 1, -1, 0, 0, 0, 1, 0, 0}]};
\end{verbatim}
\end{verse}
the boundaries for the corresponding differential equation can be computed with
\begin{verse}
\begin{verbatim}
SectorBoundariesFromID[masterIntegrals]
\end{verbatim}
\end{verse}
provided the integrals are ordered with respect to their sector-id. \texttt{\justify SectorBoundariesFromID} then returns the sector boundaries derived from the sector-ids of the integrals in the above format
\begin{verse}
\begin{verbatim}
{{1, 1}, {2, 2}, {3, 3}, {4, 4}, {5, 6}}
\end{verbatim}
\end{verse}
While the main function of \textit{CANONICA} is \texttt{\justify RecursivelyTransformSectors}, it is in some cases useful to be able to perform only certain steps of the algorithm. For this reason, there is a hierarchy of functions available in \textit{CANONICA} allowing to break the calculation of the transformation into smaller steps. The hierarchy of these lower-level functions is illustrated in Fig. \ref{fig:hierarchy}. For more information on specific functions, see the manual notebook included in the package.

\subsection{Tests and Limitations}

\textit{CANONICA} has been successfully tested on a variety of non-trivial single- and multi-scale problems, some of which are included as examples in the package. All tests have been performed with the \texttt{\justify Mathematica} versions 10 and 11 on a Linux operating system. The limitations of \textit{CANONICA} are mostly limitations of the algorithm itself. In particular, the algorithm is limited to differential equations for which a rational transformation to a canonical form exists. However, it is well known that rational differential equations may require non-rational transformations to attain a canonical form. The following example illustrates this behavior: 
\begin{equation}
a(\epsilon,\{x\})=\left(\frac{1}{2x}+\frac{\epsilon}{x}\right)\textup{d}x,
\end{equation}
where the transformation to a canonical form is given by
\begin{equation}
T(\epsilon,\{x\})=\sqrt{x}.
\end{equation}
In this situation it is often possible to render the transformation rational with a change of coordinates. For instance, in the above example, the differential form transforms under the change of variables 
\begin{equation}
x=y^2
\end{equation}
to
\begin{equation}
a(\epsilon,\{y\})=\left(\frac{1}{y}+\frac{2\epsilon}{y}\right)\textup{d}y,
\end{equation}
which has the following rational transformation to a canonical form
\begin{equation}
T(\epsilon,\{y\})=y.
\end{equation}
While a change of coordinates can remove non-rational letters in more complicated examples as well \cite{Henn:2013woa, Henn:2014lfa, Gehrmann:2014bfa, Caola:2014lpa}, this has neither been proven to be always possible, nor is a general method to construct such coordinate changes known. In fact, the existence of such a procedure appears to be unlikely, given that the number of independent roots can largely outgrow the number of variables in a problem \cite{Bonciani:2016qxi}.

In addition to this limitation of the algorithm, the calculations are in practice limited by the size of the available memory. This imposes limits on the size of the systems of linear parameter equations. The main factors determining the sizes of these systems of linear equations are the size of the differential equation itself and the size of the ansatz. Thus, run time and memory consumption of \textit{CANONICA} are highly problem dependent. For instance, the most complicated example that is provided with the package is a two-loop double box topology depending on three dimensionless scales. It has a run time of about 20 minutes and a memory consumption of less than 8 GB.

\section{Conclusion}
\label{sec:Conclusion}
The description of the algorithm in Ref. \cite{Meyer:2016slj} has been extended in the present paper by providing details on the choice of an ansatz for both the diagonal blocks and the off-diagonal blocks. In both cases it has been shown that the irreducible denominator factors can be chosen from the set of irreducible denominator factors occurring in the differential equation. The ansatz for the off-diagonal blocks has been further restricted by proving upper bounds on the powers of the denominator factors in the solution. 

Furthermore, canonical forms have been proven to be unique up to constant transformations, which allowed to attribute the occurrence of non-linear equations in the parameters to the freedom in the choice of this constant transformation. By fixing this freedom in a specific way, it has been argued that only linear equations need to be solved in the course of applying the algorithm.

The main focus of this publication has been the presentation of the \texttt{\justify Mathematica} package \textit{CANONICA}, which implements the aforementioned algorithm and allows to compute rational transformations of differential equations into canonical form. This represents the first publicly available implementation of an algorithm applicable to problems depending on multiple scales. \textit{CANONICA} has been successfully tested on a number of state of the art multi-scale problems, including previously unknown integral topologies. \textit{CANONICA} may thus provide a valuable contribution to the ongoing efforts to automatize multi-loop calculations.

\section*{Acknowledgments}
The author would like to thank Peter Uwer for useful discussions and comments on the manuscript. This research was supported by the German Research Foundation (DFG) via the Research Training Group 1504.

\newpage
\appendix

\section{List of functions provided by \textit{CANONICA}} 
\label{App:ListFunctions}
\begin{description}
\item[CalculateDlogForm:]~\\\texttt{\justify CalculateDlogForm[a, invariants, alphabet]} returns a list of matrices of the same dimensions as \texttt{\justify a}, where each matrix is the matrix-residue of one of the letters. The ordering is the same as the one in \texttt{\justify alphabet}. Returns \texttt{\justify False} if \texttt{\justify a} cannot be cast in a dlog-form with the given \texttt{\justify alphabet}.
\item[CalculateNexta:]~\\\texttt{\justify CalculateNexta[aFull, invariants, sectorBoundaries, trafoPrevious, aPrevious]} applies \texttt{\justify trafoPrevious} to \texttt{\justify aFull} and returns the differential equation of the next sector. \texttt{\justify aPrevious} is used to recycle the transformation of lower sectors.
\item[CalculateNextSubsectorD:]~\\\texttt{\justify CalculateNextSubsectorD[a, invariants, sectorBoundaries, previousD]} computes the $D_k$ of the next sector, prepends it to \texttt{\justify previousD} and returns the result. The ansatz to be used can be specified with the optional argument \texttt{\justify userProvidedAnsatz}. If no ansatz is provided, an ansatz is generated automatically. The size of the automatically generated ansatz can be controlled with the option \texttt{\justify DDeltaNumeratorDegree}.
\item[CheckDlogForm:]~\\\texttt{\justify CheckDlogForm[a, invariants, alphabet]} tests whether the differential equation \texttt{\justify a} is in dlog-form for the given \texttt{\justify alphabet}. Returns either \texttt{\justify True} or \texttt{\justify False}.
\item[CheckEpsForm:]~\\\texttt{\justify CheckEpsForm[a, invariants, alphabet]} tests whether the differential equation \texttt{\justify a} is in $\epsilon$-form with the given \texttt{\justify alphabet}. Returns either \texttt{\justify True} or \texttt{\justify False}.
\item[CheckIntegrability:]~\\\texttt{\justify CheckIntegrability[a, invariants]} tests whether \texttt{\justify a} satisfies the integrability condition $\textup{d}a-a\wedge a=0$ and returns either \texttt{\justify True} or \texttt{\justify False}.
\item[CheckSectorBoundaries:]~\\\texttt{\justify CheckSectorBoundaries[a, sectorBoundaries]} tests whether the \texttt{\justify sectorBoundaries} are compatible with \texttt{\justify a} and returns either \texttt{\justify True} or \texttt{\justify False}.
\item[ExtractDiagonalBlock:]~\\\texttt{\justify ExtractDiagonalBlock[a, boundaries]} returns the diagonal block of the differential equation \texttt{\justify a} specified by the \texttt{\justify boundaries} argument. \texttt{\justify boundaries} is expected to be of the format \texttt{\justify \{nLowest, nHighest\}}, where \texttt{\justify nLowest} and \texttt{\justify nHighest} are positive integers indicating the lowest and highest integrals of the diagonal block, respectively.
\item[ExtractIrreducibles:]~\\\texttt{\justify ExtractIrreducibles[a]} returns the irreducible denominator factors of \texttt{\justify a} that do not depend on the regulator. The option \texttt{\justify AllowEpsDependence->True} allows the irreducible factors to depend on both the invariants and the regulator.
\item[FindAnsatzSubsectorD:]~\\\texttt{\justify FindAnsatzSubsectorD[a, invariants, sectorBoundaries, previousD]} takes a differential equation \texttt{\justify a}, which is required to be in $\epsilon$-form except for the off-diagonal block of the highest sector. Needs to be provided with all previous $D_k$ in the argument \texttt{\justify previousD} and computes the ansatz $\mathcal{R}_D$ for the computation of the next $D_k$. Takes the option \texttt{\justify DDeltaNumeratorDegree} to enlarge the ansatz. For more details, see Section \ref{subsec:AnsatzOD}.
\item[FindAnsatzT:]~\\\texttt{\justify FindAnsatzT[a, invariants]} takes a differential equation \texttt{\justify a} in the \texttt{\justify invariants} and computes an ansatz $\mathcal{R}_T$ as described in Section \ref{subsec:AnsatzDB}. The ansatz can be enlarged with the options \texttt{\justify TDeltaNumeratorDegree} and \texttt{\justify TDeltaDenominatorDegree}.
\item[FindConstantNormalization:]~\\\texttt{\justify FindConstantNormalization[invariants, trafoPrevious, aPrevious]} calculates a constant diagonal transformation to minimize the number of prime factors present in the matrix-residues. The transformation is composed with \texttt{\justify trafoPrevious} and returned together with the resulting differential equation.
\item[FindEpsDependentNormalization:]~\\\texttt{\justify FindEpsDependentNormalization[a, invariants]} calculates a diagonal transformation depending only on the dimensional regulator in order to attempt to minimize the number of orders that need to be calculated in a subsequent determination of the transformation to a canonical form. Returns the transformation together with the resulting differential equation.
\item[RecursivelyTransformSectors:]~\\\texttt{\justify RecursivelyTransformSectors[aFull, invariants, sectorBoundaries, \{nSecStart, nSecStop\}]} calculates a rational transformation of \texttt{\justify aFull} to a canonical form in a recursion over the sectors of the differential equation, which have to be specified by \texttt{\justify sectorBoundaries}. The arguments \texttt{\justify nSecStart} and \texttt{\justify nSecStop} set the first and the last sector to be computed, respectively. If \texttt{\justify nSecStart} is greater than one, the result of the calculation for the sectors lower than \texttt{\justify nSecStart} needs to be provided in the additional arguments \texttt{\justify trafoPrevious} and \texttt{\justify aPrevious}. \texttt{\justify RecursivelyTransformSectors} returns the transformation of \texttt{\justify aFull} to a canonical form for the sectors up to \texttt{\justify nSecStop} and the resulting differential equation. The ansaetze for the individual blocks are generated automatically. The sizes of the ansaetze for the diagonal blocks can be controlled with the options \texttt{\justify TDeltaNumeratorDegree} and \texttt{\justify TDeltaDenominatorDegree}. Similarly, the sizes of the ansaetze for the off-diagonal blocks are controlled by the option \texttt{\justify DDeltaNumeratorDegree}.
\item[SectorBoundariesFromDE:]~\\\texttt{\justify SectorBoundariesFromDE[a]} returns the most fine grained sector boundaries compatible with \texttt{\justify a}.
\item[SectorBoundariesFromID:]~\\\texttt{\justify SectorBoundariesFromID[masterIntegrals]} takes a list of \texttt{\justify masterIntegrals}, which need to be ordered by their sector-id and returns the sector boundaries computed from the sector-ids.
\item[TransformDE:]~\\\texttt{\justify TransformDE[a, invariants, t]} applies the transformation \texttt{\justify t} to the differential equation \texttt{\justify a}. Returns $a^\prime=t^{-1}at-t^{-1}\textup{d}t$. The option \texttt{\justify SimplifyResult->False} deactivates the simplification of the result.
\item[TransformDiagonalBlock:]~\\\texttt{\justify TransformDiagonalBlock[a, invariants]} calculates a rational transformation to transform \texttt{\justify a} into canonical form and returns the transformation together with the resulting differential equation. With the optional argument \texttt{\justify userProvidedAnsatz}, the user can specify the ansatz to be used. If no ansatz is provided, an ansatz is generated automatically. The size of the automatically generated ansatz can be controlled with the options \texttt{\justify TDeltaNumeratorDegree} and \texttt{\justify TDeltaDenominatorDegree}.
\item[TransformDlogToEpsForm:]~\\\texttt{\justify TransformDlogToEpsForm[invariants, sectorBoundaries, trafoPrevious, aPrevious]} computes a transformation depending only on the regulator in order to transform \texttt{\justify aPrevious} from dlog-form into canonical form (cf. Ref. \cite{Lee:2014ioa}). The transformation is composed with \texttt{\justify trafoPrevious} and returned together with the resulting differential equation. Per default, the transformation is demanded to be in a block-triangular form induced by \texttt{\justify sectorBoundaries}. This condition can be dropped with the option \texttt{\justify EnforceBlockTriangular->False}.
\item[TransformNextDiagonalBlock:]~\\\texttt{\justify TransformNextDiagonalBlock[aFull, invariants, sectorBoundaries, trafoPrevious, aPrevious]} calls \texttt{\justify TransformDiagonalBlock} to compute the transformation of the next diagonal block into canonical form and composes it with \texttt{\justify trafoPrevious}. Returns the composed transformation together with the resulting differential equation. With the optional argument \texttt{\justify userProvidedAnsatz}, the user can specify the ansatz to be used. If no ansatz is provided, an ansatz is generated automatically. The size of the automatically generated ansatz can be controlled with the options \texttt{\justify TDeltaNumeratorDegree} and \texttt{\justify TDeltaDenominatorDegree}.
\item[TransformNextSector:]~\\\texttt{\justify TransformNextSector[aFull, invariants, sectorBoundaries, trafoPrevious, aPrevious]} transforms the next sector into canonical form, composes the calculated transformation with \texttt{\justify trafoPrevious} and returns it together with the resulting differential equation. With the optional argument \texttt{\justify userProvidedAnsatz}, the user can specify the ansatz to be used for the diagonal block. If no ansatz is provided, an ansatz is generated automatically. The size of the automatically generated ansatz for the diagonal block can be controlled with the options \texttt{\justify TDeltaNumeratorDegree} and \texttt{\justify TDeltaDenominatorDegree}. Similarly, the sizes of the ansaetze for the off-diagonal blocks are controlled by the option \texttt{\justify DDeltaNumeratorDegree}.
\item[TransformOffDiagonalBlock:]~\\\texttt{\justify TransformOffDiagonalBlock[invariants, sectorBoundaries, trafoPrevious, aPrevious]} assumes \texttt{\justify aPrevious} to be in canonical form except for the highest sector of which only the diagonal block is assumed to be in canonical form. Computes a transformation to transform the off-diagonal block of the highest sector into dlog-form. This transformation is composed with \texttt{\justify trafoPrevious} and returned together with the resulting differential equation. Proceeds in a recursion over sectors, which can be resumed at an intermediate step by providing all previous $D_k$ (cf. Ref. \cite{Meyer:2016slj}) in the optional argument \texttt{\justify userProvidedD}. The sizes of the automatically generated ansaetze for the off-diagonal blocks are controlled by the option \texttt{\justify DDeltaNumeratorDegree}.
\end{description}

\section{List of options} 
\label{App:ListOptions}
\begin{description}
\item[AllowEpsDependence:]~\\\texttt{\justify AllowEpsDependence} is an option of \texttt{\justify ExtractIrreducibles} controlling whether irreducible factors depending on both the invariants and the regulator are returned as well. The default value is \texttt{\justify False}.
\item[DDeltaNumeratorDegree:]~\\\texttt{\justify DDeltaNumeratorDegree} is an option controlling the numerator powers of the rational functions in the ansatz used for the computation of $D$ for the transformation of off-diagonal blocks. The default value is \texttt{0}. For more details, see Section \ref{subsec:AnsatzOD}. \texttt{\justify DDeltaNumeratorDegree} is an option of the following functions: \texttt{\justify CalculateNextSubsectorD}, \texttt{\justify FindAnsatzSubsectorD}, \texttt{\justify RecursivelyTransformSectors}, \texttt{\justify TransformNextSector}, \texttt{\justify TransformOffDiagonalBlock}.
\item[EnforceBlockTriangular:]~\\\texttt{\justify EnforceBlockTriangular} is an option of \texttt{\justify TransformDlogToEpsForm} controlling whether the resulting transformation is demanded to be in the block-triangular form induced by the \texttt{\justify sectorBoundaries} argument. The default value is \texttt{\justify True}.
\item[FinalConstantNormalization:]~\\\texttt{\justify FinalConstantNormalization} is an option of \texttt{\justify RecursivelyTransformSectors} controlling whether \texttt{\justify FindConstantNormalization} is invoked after all sectors have been transformed into canonical form in order to simplify the resulting canonical form. The default value is \texttt{\justify False}.
\item[PreRescale:]~\\\texttt{\justify PreRescale} is an option of \texttt{\justify TransformDiagonalBlock} controlling whether \texttt{\justify FindEpsDependentNormalization} is called prior to the main computation in order to attempt to minimize the number of orders that need to be calculated in a subsequent determination of the transformation to a canonical form. The default value is \texttt{\justify True}.
\item[SimplifyResult:]~\\\texttt{\justify SimplifyResult} is an option of \texttt{\justify TransformDE} controlling whether the resulting differential equation is simplified. The default value is \texttt{\justify True}.
\item[TDeltaDenominatorDegree:]~\\\texttt{\justify TDeltaDenominatorDegree} is an option controlling the denominator powers of the rational functions in the ansatz used for the computation of the transformation of diagonal blocks. The default value is \texttt{0}. For more details, see Section \ref{subsec:AnsatzDB}. \texttt{\justify TDeltaDenominatorDegree} is an option of the following functions: \texttt{\justify FindAnsatzT}, \texttt{\justify RecursivelyTransformSectors}, \texttt{\justify TransformNextDiagonalBlock}, \texttt{\justify TransformNextSector}.
\item[TDeltaNumeratorDegree:]~\\\texttt{\justify TDeltaNumeratorDegree} is an option controlling the numerator powers of the rational functions in the ansatz used for the computation of the transformation of diagonal blocks. The default value is \texttt{0}. For more details, see Section \ref{subsec:AnsatzDB}. \texttt{\justify TDeltaNumeratorDegree} is an option of the following functions: \texttt{\justify FindAnsatzT}, \texttt{\justify RecursivelyTransformSectors}, \texttt{\justify TransformNextDiagonalBlock}, \texttt{\justify TransformNextSector}.
\item[VerbosityLevel:]~\\\texttt{\justify VerbosityLevel} is an option controlling the verbosity of several main functions. Takes integer values from \texttt{0} to \texttt{12} with a value of \texttt{12} resulting in the most detailed output about the current state of the computation and a value of \texttt{0} suppressing all output but warnings about inconsistent inputs. The default value is \texttt{10}. The following functions accept the \texttt{\justify VerbosityLevel} option: \texttt{\justify CalculateNextSubsectorD}, \texttt{\justify FindConstantNormalization}, \texttt{\justify RecursivelyTransformSectors}, \texttt{\justify TransformDiagonalBlock}, \texttt{\justify TransformDlogToEpsForm}, \texttt{\justify TransformNextDiagonalBlock}, \texttt{\justify TransformNextSector}, \texttt{\justify TransformOffDiagonalBlock}.
\end{description}
\section{List of global variables and protected symbols} 
\label{App:GlobalVars}
\begin{description}
\item[\$ComputeParallel:]~\\\texttt{\justify \$ComputeParallel} is a global variable that needs to be set to \texttt{\justify True} to enable parallel computations. The number of kernels to be used is controlled by \texttt{\justify \$NParallelKernels}.
\item[\$NParallelKernels:]~\\\texttt{\justify \$NParallelKernels} is a global variable setting the number of parallel kernels to be used. \texttt{\justify \$NParallelKernels} has no effect if \texttt{\justify \$ComputeParallel} is not set to \texttt{\justify True}. If \texttt{\justify \$ComputeParallel} is \texttt{\justify True} and \texttt{\justify \$NParallelKernels} is not assigned a value, then all available kernels are used for the computation.
\item[eps:]~\\\texttt{\justify eps} is a protected symbol representing the dimensional regulator.
\end{description}




\bibliographystyle{elsarticle-num}
\bibliography{../../literature.bib}







\end{document}